\documentclass{elsart}

\usepackage{natbib}

\usepackage{epsfig}

\usepackage{amssymb}

\begin{document}

\begin{frontmatter}

\title{Stellar $\beta^{\pm}$ decay rates
of iron isotopes and its implications in astrophysics}

\author{Jameel-Un Nabi\thanksref{footnote2}}
\address{Faculty of Engineering Sciences, GIK Institute of Engineering
Sciences and Technology, Topi 23640, NWFP, Pakistan}
\thanks[footnote2]{Current
Address: The Abdus Salam ICTP, Strada Costiera 11, 34014, Trieste, Italy}
\ead{jameel@giki.edu.pk, jnabi@ictp.it\\ Telephone: 0092-938-271858\\Fax:
0092-938-271890}

\begin{abstract}

$\beta$-decay and positron decay are believed to play a
consequential role  during the late phases of stellar evolution of a
massive star culminating in a supernova explosion. The $\beta$-decay
contributes in maintaining a respectable lepton-to-baryon ratio,
$Y_{e}$, of the core prior to collapse which results in a larger
shock energy to produce the explosion. The positron decay acts in
the opposite direction and tends to decrease the ratio. The
structure of the presupernova star is altered both by the changes in
$Y_{e}$ and the entropy of the core material. Recently the
microscopic calculation of weak-interaction mediated rates on key
isotopes of iron was introduced using the proton-neutron
quasiparticle random phase approximation (pn-QRPA) theory with
improved model parameters. Here I discuss in detail the improved
calculation of $\beta^{\pm}$ decay rates for iron isotopes
($^{54,55,56}$Fe) in stellar environment. The pn-QRPA theory allows
a microscopic "state-by-state" calculation of stellar rates as
explained later in text. Excited state Gamow-Teller distributions
are much different from ground state and a microscopic calculation
of decay rates from these excited states greatly increases the
reliability of the total decay rate calculation specially during the
late stages of stellar evolution. The reported decay rates are also
compared with earlier calculations. The positron decay rates are in
reasonable agreement with the large-scale shell model calculation.
The main finding of this work includes that the stellar
$\beta$-decay rates of $^{54,55,56}$Fe are around 3 -- 5 orders of
magnitude smaller than previously assumed and hence irrelevant for
the determination of the evolution of $Y_{e}$ during the
presupernova phase of massive stars. The current work discourages
the inclusion of $^{55,56}$Fe in the list of key stellar
$\beta$-decay nuclei as suggested by former simulation results.

\end{abstract}

\begin{keyword}
beta decay  \sep GT strength distributions \sep pn-QRPA \sep
core-collapse supernovas \sep massive stars

\PACS 21.60.Jz \sep 23.40.-s \sep 26.50.+x \sep 97.10.Cv
\end{keyword}

\end{frontmatter}

\parindent=0.5 cm

\section{Introduction}
The classical and pioneering works in the field of stellar physics
include dynamics of supernova explosion by Baade $\&$ Zwicky
\cite{Baa34}, energy production in stars and stellar evolution by
Bethe \cite{Bet39}, and, synthesis of elements in stars by Burbidge
et al. \cite{Bur57}. Since then the microphysics of supernova
explosion has come a long way. Whereas the strong interactions
(fusion reactions) are responsible for providing the fuel to the
stars empowering them throughout their life cycles, it is precisely
the weak interactions that play a decisive role in determining both
the presupernova stellar structure and the nucleosynthesis. The weak
interaction reactions lead to the initiation of the gravitational
collapse of the core of a massive star triggering a supernova
explosion, control the lepton-to-baryon fraction of the core
throughout the course of stellar evolution, play a key role in
neutronisation of the core material, and, affect the formation of
heavy elements above iron via the r-process at the final stage of
the supernova explosion. Much to the advantage of the
astrophysicists, the temperature during the late phases of stellar
evolution is high enough for the matter composition to be given by
nuclear statistical equilibrium. This means that one can get away
with the need of reaction networks for the strong and
electromagnetic interactions and the composition of the matter is
given by the Saha equation.

Weak interactions in presupernova stars are known to be dominated by
allowed Fermi (vector-type) and Gamow-Teller (axial-vector type)
transitions. The calculation of weak-interaction rates is very
sensitive to the distribution of the GT$_{\pm}$ strength function.
In the GT$_{+}$ strength a proton is changed into a neutron whereas
the GT$_{-}$ strength is responsible for transforming a neutron into
a proton. It was Fuller, Fowler, and Newman (FFN) \cite{Ful80} who
first performed an extensive calculation of stellar weak rates
including the capture rates, decay rates, neutrino energy loss rates
and gamma heating rates for a wide density and temperature domain.
They performed their detailed calculation for 226 nuclei in the mass
range $21 \leq A \leq 60$. The authors recognized the key role
played by the GT giant resonance and noted that measured decay rates
exploited only a small fraction of the total available strength. The
centroids of the GT$_{\pm}$ distribution functions determine the
effective energy of the capture and decay reactions. FFN estimated
the GT contributions to the rates by a parametrization based on the
independent particle model. Aufderheide et al. \cite{Auf94} later
extended the FFN work for heavier nuclei with A $>$ 60 and took into
consideration the quenching of the GT strength neglected by FFN.
Authors in Ref. \cite{Auf94} also stressed on the importance of
$\beta$-decay rates in the iron core prior to the collapse. They
tabulated the 71 top $\beta$-decay nuclei averaged throughout the
stellar trajectory for $0.40 \leq Y_{e} \leq 0.5$ (see Table 26 of
Ref. \cite{Auf94}). The measured data from various $(p,n)$ and
$(n,p)$ experiments later revealed the misplacement of the GT
centroids adopted in the parameterizations of FFN and subsequently
used in the calculation of weak rates by Ref. \cite{Auf94}. Since
then theoretical efforts were concentrated on the microscopic
calculations of weak-interaction mediated rates of iron-regime
nuclide. Two such widely used models are the large-scale shell model
(LSSM)(e.g. \cite{Lan00}) and the proton-neutron quasiparticle
random phase approximation (pn-QRPA) theory  (e.g. \cite{Nab04}).

The pn-QRPA theory is an efficient way to generate GT strength
distributions which constitute a primary and nontrivial contribution
to the weak-interaction mediated rates among iron-regime nuclide.
The shell model calculations perform a detailed study of the nuclear
spectroscopy, however, the pn-QRPA model has two important
advantages. It can handle any arbitrarily heavy system of nucleons
since the calculation is performed in a luxurious model space of up
to 7 major oscillator shells. The second advantage is even more
important for the calculation of weak-interaction rates in stellar
matter. Because of the high temperatures prevailing during the
presupernova and supernova phases of a massive star, there is a
reasonable probability of occupation of parent excited states and
the total weak interaction rates have a finite contribution form
these excited states. Thus, in calculating a stellar rate, one must
know the GT strength distributions of the excited states of the
parent nucleus. As experimental information about excited state
strength functions seems inaccessible, Aufderheide \cite{Auf91}
stressed much earlier the need to probe these strength functions
theoretically. The pn-QRPA  model calculates the GT strength
distribution strengths of \textit{all} excited states of parent
nucleus in a microscopic fashion and this feature of the pn-QRPA
model greatly enhances the reliability of the calculated rates in
stellar matter. These excited states are like resonances having
finite band width and contributions from many discrete states
calculated microscopically within the pn-QRPA framework. The
construction of these excited states and calculation of the relevant
nuclear matrix elements will be shown in the next section. In this
sense the pn-QRPA model allows a fully microscopic "state-by-state"
calculation of stellar weak rates. The calculation of stellar weak
rates on iron-regime nuclei are sensitive to both the weak
low-energy and strong high-energy ground and parent excited state GT
strength distributions. The low energy part is more important during
the earlier phase and the high energy part becomes important during
the late phases of stellar evolution \cite{Auf91}. Previous pn-QRPA
calculations have shown that for certain nuclei the excited state
rates can command the total weak rate (e.g. \cite{Nab07}). This
clearly endorses the contribution of excited states in the
calculation of total weak rates. Other calculations revert to
approximations like the so-called Brink's hypothesis (in the
electron capture direction) and back-resonances (in the
$\beta$-decay direction). Brink's hypothesis states that GT strength
distribution on excited states is \textit{identical} to that from
ground state, shifted \textit{only} by the excitation energy of the
state. GT back resonances are the states reached by the strong GT
transitions in the inverse process (electron capture) built on
ground and excited states.

Nabi and Klapdor-Kleingrothaus \cite{Nab99} first reported the
calculation of weak interaction rates for 709 nuclei with A = 18 to
100 in stellar environment using the pn-QRPA theory. These included
capture rates, decay rates, gamma heating rates, neutrino energy
loss rates, probabilities of beta-delayed particle emissions and
energy rate of these particle emissions. The authors then presented
a detailed calculation of stellar weak interaction rates over a wide
range of temperature and density scale for sd- \cite{Nab99a} and
fp/fpg-shell nuclei \cite{Nab04}. These also included the weak
interaction rates for nuclei with A = 40 to 44 (not yet calculated
by shell model). Since then these calculations were further refined
with use of more efficient algorithms, computing power,
incorporation of latest data from mass compilations and experimental
values, and fine-tuning of model parameters both in the sd- shell
\cite{Nab07a, Nab08, Nab08a} and fp-shell \cite{Nab05, Nab07,
Nab07b, Nab07c, Nab08b, Nab08c, Nab08d} region. All theoretical
calculations of stellar weak interactions have inherent
uncertainties. The uncertainties associated with the pn-QRPA model
were discussed in detail in Ref. \cite{Nab08c}. The reliability of
the pn-QRPA calculation was discussed in length by Nabi and
Klapdor-Kleingrothaus \cite{Nab04}. There the authors compared the
measured data (half lives and B(GT$_{\pm}$) strength) of thousands
of nuclide with the pn-QRPA calculation and got fairly good
comparison. Earlier half-lives of $\beta^{-}$ decays were calculated
systematically for about 6000 neutron-rich nuclei between the beta
stability line and the neutron drip line using the pn-QRPA model
\cite{Sta90}. Similarly half-lives for $\beta^{+}$/EC (electron
capture) decays for neutron-deficient nuclei with atomic numbers Z =
10 - 108 were calculated up to the proton drip line for more than
2000 nuclei using the same model \cite{Hir93}. These microscopic
calculations gave a remarkably good agreement with the then existing
experimental data (within a factor of two for more than 90$\%$
(73$\%$) of nuclei with experimental half-lives shorter than 1 s for
$\beta^{-}$ ($\beta^{+}$/EC) decays). Most nuclei of interest of
astrophysical importance are the ones far from stability and one has
to rely on theoretical models to estimate their beta decay
properties. The accuracy of the pn-QRPA model increases with
increasing distance from the $\beta$-stability line (shorter
half-lives) \cite{Sta90, Hir93}. This is a promising feature with
respect to the prediction of experimentally unknown half-lives
(specially those present in the stellar interior), implying that the
predictions are made on the basis of a realistic physical model.

The isotopes of iron, $^{54,55,56}$Fe, are mainly responsible for
decreasing the electron-to-baryon ratio during the oxygen and
silicon burning phases of massive stars through electron capture and
positron decay processes. The electron captures on these iron
isotopes are the dominant process. Nevertheless, the $\beta$-decay
rates for these isotopes of iron are also argued to be relevant
during the presupernova evolution of massive stars in literature.
Because of their astrophysical importance $^{55,56}$Fe were included
in the list of key $\beta$-decay nuclei that have a significant
impact on the presupernova evolution of massive stars after core
silicon burning for $0.47 \leq Y_{e} \leq 0.49$ compiled by
Aufderheide and collaborators (see Tables 19, 20 and 26 of Ref.
\cite{Auf94}). Later Heger and collaborators \cite{Heg01} studied
the presupernova evolution of massive stars (of masses $15M_{\odot},
25M_{\odot},$ and $ 40M_{\odot}$) and found $^{55}$Fe in the list of
top five nuclei that increase $Y_{e}$ via $\beta$-decay and positron
capture during the silicon burning phases of these massive stars.
The authors employed the LSSM  $\beta$-decay rates in their
simulation code. The calculations of GT$_{\pm}$ strength
distributions and stellar weak rates for these isotopes of iron were
introduced earlier using the pn-QRPA theory with improved model
parameters \cite{Nab09a}. There the author was able to reproduce
fairly well the experimental centroids and the total strength
distributions in both directions for the even-even iron isotopes,
$^{54,56}$Fe where measurements were available.  In this paper I
discuss in detail the calculation of $\beta^{\pm}$ decay rates of
$^{54,55,56}$Fe in stellar environment and its astrophysical
implications. The main finding of this work is that $\beta$-decay
rates on $^{54,55,56}$Fe are around 3 -- 5 orders of magnitude
smaller than previously assumed and are irrelevant for the
determination of the evolution of $Y_{e}$ during the presupernova
phases of massive stars.

In the following section I briefly describe the theoretical
formalism used to calculate the stellar electron and positron decay
rates. The stellar $\beta^{\pm}$-decay rates of $^{54,55,56}$Fe are
presented in Section 3. Here I also compare the pn-QRPA decay rates
with earlier calculations.  Summary and conclusions are finally
presented in Section 4.

\section{Model Description}
The calculation may be divided into three main steps. A Bogoliubov
transformation was used to define quasiparticle states in terms of
nucleon states and then the RPA equation was solved in the basis of
proton-neutron quasiparticle pairs. The last step was to calculate
the decay rates in stellar interior. The Hamiltonian for the
calculation was of the form
\begin{equation}
H^{QRPA} =H^{sp} +V^{pair} +V_{GT}^{ph} +V_{GT}^{pp}.
\end{equation}
The first step was approximated by a Nilsson + BCS calculation.
Single-particle energies and wave functions were calculated in the
Nilsson model, which takes into account nuclear deformation
\cite{Nil55}. Pairing was treated in the BCS approximation. In the
second step regarding the RPA calculation, both particle-hole (ph)
and particle-particle (pp) GT forces were employed. The GT forces
are appropriate for calculation of GT strength functions, since the
transition amplitudes are closely connected to the forces. The
particle-particle interaction, first considered by Cha \cite{Cha83},
has usually been neglected in $\beta^{-}$ decay (e.g. \cite{Sta90})
but is of decisive importance in $\beta^{+}$ decay (e.g.
\cite{Sta90a}). Both the particle-hole and particle-particle
interaction can be given a separable form. The interactions were
characterized by two interaction constants: $\chi$ (for
particle-hole interaction) and $\kappa$ (for particle-particle
interaction). In this work, the values of $\chi$ and $\kappa$ were
taken as 0.15 MeV and 0.07 MeV, respectively. The value of the
strength parameters was determined by a fit to the measured data (GT
strengths and centroids) available for the isotopic chain. Other
parameters required for the calculation of weak rates are the
pairing gaps, the nuclear deformations and the Q-value of the
nuclear reactions. I applied the traditional choice of $\Delta _{p}
=\Delta _{n} =12/\sqrt{A} (MeV)$ in this project \cite{Hir91}. The
deformation parameter was recently argued as an important parameter
for QRPA calculations at par with the pairing parameter by Stetcu
and Johnson \cite{Ste04}. As such rather than using deformations
from some theoretical mass model (as used in earlier calculations of
pn-QRPA weak rates \cite{Nab99a, Nab04}) the experimentally adopted
values of the deformation parameters for $^{54,56}$Fe, extracted by
relating the measured energy of the first $2^{+}$ excited state with
the quadrupole deformation, were taken from Raman et al.
\cite{Ram87}. For the case of $^{55}$Fe (where such measurement
lacks) the deformation of the nucleus was calculated as
\begin{equation}
\delta = \frac{125(Q_{2})}{1.44 (Z) (A)^{2/3}},
\end{equation}
where $Z$ and $A$ are the atomic and mass numbers, respectively, and
$Q_{2}$ is the electric quadrupole moment taken from Ref.
\cite{Moe81}. Q-values were taken from the recent mass compilation
of Audi et al. \cite{Aud03}.

The incorporation of measured deformations for $^{54,56}$Fe and a
smart choice of strength parameters led to an improvement of the
calculated GT$_{\pm}$ distributions compared to the measured ones.
Excited state calculation of GT$_{\pm}$ strengths was performed for
a total of 246 parent states in $^{54}$Fe, 297 states in $^{55}$Fe
and 266 states in $^{56}$Fe, covering excitation energies in the
vicinity of 15 MeV (as explained earlier). For each parent excited
state, transitions were calculated to 150 daughter excited states.
Daughter excitation energies up to around 20 MeV were considered in
the calculation. The band widths of energy states were chosen
according to the density of states of the concerned nucleus. In this
way contribution from all excited states was incorporated in the
calculation. A large model space assists in reproducing low-lying
spectrum and higher excitations \cite{Hax90}. The use of a separable
interaction assisted in the incorporation of a luxurious model space
of up to 7 major oscillator shells which in turn made possible to
consider these many excited states in both parent and daughter
nuclei. In order to further increase the reliability of the
calculated rates experimental data were incorporated in the
calculation wherever possible. The calculated excitation energies
were replaced with measured levels when they were within 0.5 MeV of
each other. Missing measured states were inserted and inverse
transitions (along with their log$ft$ values) were also taken into
account. No theoretical levels were replaced with the experimental
ones beyond the excitation energy for which experimental
compilations had no definite spin and/or parity.

The beta decay (bd) and positron decay (pd) rates of a transition
from the $i^{th}$ state of the parent to the $j^{th}$ state of the
daughter nucleus are given by
\begin{equation}
\lambda ^{^{bd(pd)} } _{ij} =\left[\frac{\ln 2}{D}
\right]\left[B(F)_{ij} +\left({\raise0.7ex\hbox{$ g_{A}
$}\!\mathord{\left/ {\vphantom {g_{A}  g_{V} }} \right.
\kern-\nulldelimiterspace}\!\lower0.7ex\hbox{$ g_{V}  $}}
\right)^{2}_{eff} B(GT)_{ij} \right]\left[f_{ij}^{bd(pd)} (T,\rho
,E_{f} )\right]. \label{phase space}
\end{equation}
The value of D was taken to be 6295s \cite{Yos88}. B(F) and B(GT)
are reduced transition probabilities of the Fermi and ~Gamow-Teller
(GT) transitions, respectively,
\begin{equation}
B(F)_{ij} = \frac{1}{2J_{i}+1} \mid<j \parallel \sum_{k}t_{\pm}^{k}
\parallel i> \mid ^{2}.
\end{equation}
\begin{equation}
B(GT)_{ij} = \frac{1}{2J_{i}+1} \mid <j \parallel
\sum_{k}t_{\pm}^{k}\vec{\sigma}^{k} \parallel i> \mid ^{2}.
\end{equation}
Here $\vec{\sigma}^{k}$ is the spin operator and $t_{\pm}^{k}$
stands for the isospin raising and lowering operator. Details of the
calculation of reduced transition probabilities can be found in Ref.
\cite{Nab99a}. The effective ratio of axial and vector coupling
constants, $(g_{A}/g_{V})_{eff}$, which takes into account the
observed quenching of the GT strength was taken to be (from Ref.
\cite{Gaa83}):
\begin{equation}
\left (\frac{g_{A}}{g_{V}} \right)^{2}_{eff} = 0.60 \left
(\frac{g_{A}}{g_{V}} \right)^{2}_{bare},
\end{equation}
with $(g_{A}/g_{V})_{bare}$ = -1.254 \cite{Rod06}. Interestingly,
Vetterli and collaborators \cite{Vet89} and R\"{o}nnqvist et al.
\cite{Roe93} predicted the same quenching factor of 0.6 for the RPA
calculation in the case of $^{54}$Fe when comparing their measured
strengths to RPA calculation.

The $f_{ij}^{bd(pd)}$ are the phase space integrals and are
functions of stellar temperature ($T$), density ($\rho$) and Fermi
energy ($E_{f}$) of the electrons. They are explicitly given by
\begin{equation}
f_{ij}^{bd} \, =\, \int _{1 }^{w_{m}}w\sqrt{w^{2} -1} (w_{m} \,
 -\, w)^{3} F(+ Z,w)(1- G_{-}) dw,
 \label{bd}
\end{equation}
and by
\begin{equation}
f_{ij}^{pd} \, =\, \int _{1 }^{w_{m}}w\sqrt{w^{2} -1} (w_{m} \,
 -\, w)^{3} F(- Z,w)(1- G_{+}) dw,
 \label{pd}
\end{equation}
In Eqs. ~(\ref{bd}) and ~(\ref{pd}), $w$ is the total energy of the
electron including its rest mass. $w_{m}$ is the total $\beta$-decay
energy,
\begin{equation}
w_{m} = m_{p}-m_{d}+E_{i}-E_{j},
\end{equation}
where $m_{p}$ and $E_{i}$ are masses and excitation energies of the
parent nucleus, and $m_{d}$ and $E_{j}$ of the daughter nucleus,
respectively. F($ \pm$ Z,w) are the Fermi functions and were
calculated according to the procedure adopted by Gove and Martin
\cite{Gov71}. G$_{\pm}$ is the Fermi-Dirac distribution function for
positrons (electrons).
\begin{equation}
G_{+} =\left[\exp \left(\frac{E+2+E_{f} }{kT}\right)+1\right]^{-1},
\end{equation}
\begin{equation}
 G_{-} =\left[\exp \left(\frac{E-E_{f} }{kT}
 \right)+1\right]^{-1},
\end{equation}
here $E$ is the kinetic energy of the electrons and $k$ is the
Boltzmann constant.

The total decay rate per unit time per nucleus is finally given by
\begin{equation}
\lambda^{bd(pd)} =\sum _{ij}P_{i} \lambda _{ij}^{bd(pd)},
\end{equation}
where $P_{i}$ is the probability of occupation of parent excited
states and follows the normal Boltzmann distribution. After the
calculation of all partial rates for the transition $i \rightarrow
j$ the summation was carried out over all initial and final states
until satisfactory convergence was achieved in the rate calculation.

Stellar decay rates are fragile functions of the available phase
space, ($ Q_{\beta} +E_{i} - E_{j}$). Under terrestrial conditions,
$\beta$-decay of $^{54,55,56}$Fe is not possible ($Q_{\beta}$ =
-8.2430 MeV, -3.4520 MeV and -4.5660 MeV, for $^{54}$Fe, $^{55}$Fe
and $^{56}$Fe, respectively \cite{Aud03}). However, for stellar
conditions the phase space can become positive depending on the
calculated energy eigenvalues of the underlying theoretical model.
Construction of parent and daughter excited states and calculation
of nuclear matrix elements in the pn-QRPA model is treated
separately in the next section.

\subsection{Calculation of excited states and nuclear matrix elements}
The excited states in the pn-QRPA model can be constructed as
phonon-correlated multi-quasi-particle states. The RPA is formulated
for excitations from the $J^{\pi} = 0^{+}$ ground state of an
even-even nucleus. The model extended to include the quasiparticle
transition degrees of freedom yields decay half-lives of odd-mass
and odd-odd parent nuclei with the same quality of agreement with
experiment as for even-even nuclei (where only QRPA phonons
contribute to the decays) \cite{Mut92}. For  the odd-A nucleus,
$^{55}$Fe, the ground state can be expressed as a one-quasiparticle
state, in which the odd quasiparticle (q.p) occupies the single-q.p.
orbit of the smallest energy.  Then there exits two different type
of transitions: phonon transitions with the odd neutron acting only
as a spectator and transition of the odd neutron itself. For the
later case, phonon correlations were introduced to one-quasiparticle
states in first-order perturbation \cite{Mut89}. The transition
amplitudes between the multi-quasi-particle states can be reduced to
those of single-quasi-particle states as shown below.

Excited states of even-even nucleus are obtained by
one-proton (or one-neutron) excitations. They are described, in the
quasiparticle (q.p.) picture, by adding two-proton (two-neutron)
q.p.'s to the ground state \cite{Mut92}. Excited states of
$^{54,56}$Fe are two-proton q.p. states and two-neutron q.p. states.
Transitions from these initial states are possible to final
proton-neutron q.p. pair states in the odd-odd daughter nucleus. The
transition amplitudes and their reduction to correlated ($c$)
one-q.p. states are given by
\begin{eqnarray}
<p^{f}n_{c}^f \mid t_{\pm}\sigma_{-\mu} \mid p_{1}^{i}p_{2c}^{i}> \nonumber \\
 = -\delta (p^{f},p_{2}^{i}) <n_{c}^{f} \mid t_{\pm}\sigma_{-\mu} \mid p_{1c}^{i}>
+\delta (p^{f},p_{1}^{i}) <n_{c}^{f} \mid t_{\pm}\sigma_{-\mu} \mid
p_{2c}^{i}>
 \label{first}
\end{eqnarray}
\begin{eqnarray}
<p^{f}n_{c}^f \mid t_{\pm}\sigma_{\mu} \mid n_{1}^{i}n_{2c}^{i}> \nonumber \\
 = +\delta (n^{f},n_{2}^{i}) <p_{c}^{f} \mid t_{\pm}\sigma_{\mu} \mid n_{1c}^{i}>
-\delta (n^{f},n_{1}^{i}) <p_{c}^{f} \mid t_{\pm}\sigma_{\mu} \mid
n_{2c}^{i}>
\end{eqnarray}
where $\mu$ = -1, 0, 1, are the spherical components of the spin
operator.

Four-proton (four-neutron) q.p. states and higher q.p. states were not considered in the model for the construction of excited states 
of $^{54,56}$Fe. Higher q.p. states may affect the total decay rates specially at parent excitation energies well in excess of 5 MeV. However 
the higher q.p. states were not considered in this calculation for consistency reasons. The model used for the calculation of weak-interaction
rates was briefly introduced in Ref. \cite{Nab09a}. This model, along with the recipe to construct excited states as described in Ref. \cite{Mut92}, was used to calculate electron and positron capture rates; electron and positron decay rates; and finally the neutrino and anti-neutrino cooling rates due to $^{54,55,56}$Fe. It was reported in Ref. \cite{Nab09a} that the calculated electron capture rates and neutrino cooling rates, using this model,  was up to a factor four bigger than the corresponding LSSM rates \cite{Lan00} for the even-even isotopes of iron. Using the same model the calculated $\beta$-decay rates were found to be 3-5 orders of magnitude smaller \cite{Nab09a} than the corresponding LSSM decay rates. The total decay rates are commanded mainly by the GT forces operating in the particle-particle and particle-hole channels. These forces were characterized by the interaction constants $\chi$ and $\kappa$ as described previously. Work is currently in progress to include higher q.p. states in the construction of parent excited states to account for possible complicated configurations expected at high excitation energies. Any effect on the calculated $\beta$-decay rates due to inclusion of higher q.p. states would be reported in future.

When a nucleus has an odd nucleon (a proton and/or a neutron), some low-lying states are obtained by lifting the q.p. in
the orbit of the smallest energy to higher-lying orbits \cite{Mut92}. For $^{55}$Fe nucleus, the excited states can be constructed\\
(1) by lifting the odd neutron from ground state to excited states (one-q.p. state),\\
(2) by three-neutron states, corresponding to excitation of a neutron (three-q.p. states), or,\\
(3) by one-neutron two-proton states, corresponding to excitation of
a proton (three-q.p. states).

The formulae for multi-q.p. transitions and their reduction to
correlated ($c$) one-q.p. states are given by,
\begin{eqnarray}
<p_{1}^{f}n_{1}^{f}n_{2c}^{f} \mid t_{\pm}\sigma_{\mu} \mid n_{1}^{i}n_{2}^{i}n_{3c}^{i}> \nonumber \\
 = \delta (n_{1}^{f},n_{2}^{i}) \delta (n_{2}^{f},n_{3}^{i}) <p_{1c}^{f} \mid t_{\pm}\sigma_{\mu} \mid n_{1c}^{i}>
 - \delta (n_{1}^{f},n_{1}^{i}) \delta (n_{2}^{f},n_{3}^{i}) \nonumber\\
 <p_{1c}^{f} \mid t_{\pm}\sigma_{\mu} \mid n_{2c}^{i}>
 + \delta (n_{1}^{f},n_{1}^{i}) \delta (n_{2}^{f},n_{2}^{i}) <p_{1c}^{f} \mid t_{\pm}\sigma_{\mu} \mid n_{3c}^{i}>
\end{eqnarray}
\begin{eqnarray}
<p_{1}^{f}n_{1}^{f}n_{2c}^{f} \mid t_{\pm}\sigma_{-\mu} \mid p_{1}^{i}p_{2}^{i}n_{1c}^{i}> \nonumber\\
 = \delta (p_{1}^{f},p_{2}^{i})[ \delta (n_{1}^{f},n_{1}^{i}) <n_{2c}^{f} \mid t_{\pm}\sigma_{-\mu} \mid p_{1c}^{i}>
 - \delta (n_{2}^{f},n_{1}^{i}) \nonumber\\
<n_{1c}^{f} \mid t_{\pm}\sigma_{-\mu} \mid p_{1c}^{i}>]
 -\delta (p_{1}^{f},p_{1}^{i})[ \delta (n_{1}^{f},n_{1}^{i}) <n_{2c}^{f} \mid t_{\pm}\sigma_{-\mu} \mid
 p_{2c}^{i}> \nonumber\\
  - \delta (n_{2}^{f},n_{1}^{i}) <n_{1c}^{f} \mid t_{\pm}\sigma_{-\mu} \mid
  p_{2c}^{i}>]
\end{eqnarray}
\begin{eqnarray}
<p_{1}^{f}p_{2}^{f}p_{3c}^{f} \mid t_{\pm}\sigma_{\mu} \mid p_{1}^{i}p_{2}^{i}n_{1c}^{i}> \nonumber\\
 = \delta (p_{2}^{f},p_{1}^{i}) \delta (p_{3}^{f},p_{2}^{i}) <p_{1c}^{f} \mid t_{\pm}\sigma_{\mu} \mid n_{1c}^{i}>
 - \delta (p_{1}^{f},p_{1}^{i}) \delta (p_{3}^{f},p_{2}^{i}) \nonumber\\
<p_{2c}^{f} \mid t_{\pm}\sigma_{\mu} \mid
 n_{1c}^{i}>+ \delta (p_{1}^{f},p_{1}^{i}) \delta (p_{2}^{f},p_{2}^{i})
<p_{3c}^{f} \mid t_{\pm}\sigma_{\mu} \mid n_{1c}^{i}>
\end{eqnarray}

Next I describe the construction of excited states of the daughter
nuclei and the calculation of the respective matrix elements in the
model.
The low-lying states of an odd-proton even-neutron nucleus (daughter of $^{55}$Fe) can be constructed,\\
(1) by exciting the odd proton from ground state (one-q.p. states),\\
(2) by three-proton states, corresponding to excitation of a proton (three-q.p. states), or,\\
(3) by one-proton two-neutron states, corresponding to excitation of
a neutron (three-q.p. states).

The multi-q.p. transitions can again be reduced to correlated ($c$)
one-q.p. states,
\begin{eqnarray}
<p_{1}^{f}p_{2}^{f}n_{1c}^{f} \mid t_{\pm}\sigma_{-\mu} \mid p_{1}^{i}p_{2}^{i}p_{3c}^{i}> \nonumber\\
 = \delta (p_{1}^{f},p_{2}^{i}) \delta (p_{2}^{f},p_{3}^{i}) <n_{1c}^{f} \mid t_{\pm}\sigma_{-\mu} \mid p_{1c}^{i}>
 - \delta (p_{1}^{f},p_{1}^{i}) \delta (p_{2}^{f},p_{3}^{i}) \nonumber\\
<n_{1c}^{f} \mid t_{\pm}\sigma_{-\mu} \mid p_{2c}^{i}> + \delta
(p_{1}^{f},p_{1}^{i}) \delta (p_{2}^{f},p_{2}^{i}) <n_{1c}^{f} \mid
t_{\pm}\sigma_{-\mu} \mid p_{3c}^{i}>
\end{eqnarray}
\begin{eqnarray}
<p_{1}^{f}p_{2}^{f}n_{1c}^{f} \mid t_{\pm}\sigma_{\mu} \mid p_{1}^{i}n_{1}^{i}n_{2c}^{i}> \nonumber\\
 = \delta (n_{1}^{f},n_{2}^{i})[ \delta (p_{1}^{f},p_{1}^{i}) <p_{2c}^{f} \mid t_{\pm}\sigma_{\mu} \mid n_{1c}^{i}>
 - \delta (p_{2}^{f},p_{1}^{i})\nonumber\\
<p_{1c}^{f} \mid t_{\pm}\sigma_{\mu} \mid n_{1c}^{i}>]  -\delta
(n_{1}^{f},n_{1}^{i})[ \delta (p_{1}^{f},p_{1}^{i})
<p_{2c}^{f} \mid t_{\pm}\sigma_{\mu} \mid n_{2c}^{i}> \nonumber \\
 - \delta (p_{2}^{f},p_{1}^{i}) <p_{1c}^{f} \mid t_{\pm}\sigma_{\mu} \mid n_{2c}^{i}>]
\end{eqnarray}
\begin{eqnarray}
<n_{1}^{f}n_{2}^{f}n_{3c}^{f} \mid t_{\pm}\sigma_{-\mu} \mid p_{1}^{i}n_{1}^{i}n_{2c}^{i}> \nonumber\\
=\delta (n_{2}^{f},n_{1}^{i}) \delta (n_{3}^{f},n_{2}^{i})
<n_{1c}^{f} \mid t_{\pm}\sigma_{-\mu} \mid p_{1c}^{i}> \mbox{}
-\delta (n_{1}^{f},n_{1}^{i}) \delta
(n_{3}^{f},n_{2}^{i})\nonumber\\
<n_{2c}^{f} \mid t_{\pm}\sigma_{-\mu} \mid p_{1c}^{i}> \mbox{}
+\delta (n_{1}^{f},n_{1}^{i}) \delta (n_{2}^{f},n_{2}^{i})
<n_{3c}^{f} \mid t_{\pm}\sigma_{-\mu} \mid p_{1c}^{i}>
\end{eqnarray}

For an odd-odd nucleus the ground state is assumed to be a
proton-neutron q.p. pair state of smallest energy. Low-lying states
in an odd-odd nucleus are expressed in the q.p. picture by
proton-neutron pair states (two-q.p. states) or by states which are
obtained by adding two-proton or two-neutron q.p.'s (four-q.p.
states) \cite{Mut92}.  Therefore states in an odd-odd nucleus
(daughter of $^{54,56}$Fe) are expressed in q.p. transformation by
two-q.p. states (proton-neutron pair states) or by four-q.p. states
(two-proton or two-neutron q.p. states). Reduction of two-q.p.
states into correlated ($c$) one-q.p. states is given as
\begin{eqnarray}
<p_{1}^{f}p_{2c}^{f} \mid t_{\pm}\sigma_{\mu} \mid p^{i}n_{c}^{i}> \nonumber\\
= \delta(p_{1}^{f},p^{i}) <p_{2c}^{f} \mid t_{\pm}\sigma_{\mu} \mid
n_{c}^{i}> - \delta(p_{2}^{f},p^{i}) <p_{1c}^{f} \mid
t_{\pm}\sigma_{\mu} \mid n_{c}^{i}>
\end{eqnarray}
\begin{eqnarray}
<n_{1}^{f}n_{2c}^{f} \mid t_{\pm}\sigma_{-\mu} \mid p^{i}n_{c}^{i}> \nonumber\\
= \delta(n_{2}^{f},n^{i}) <n_{1c}^{f} \mid t_{\pm}\sigma_{-\mu} \mid
p_{c}^{i}> - \delta(n_{1}^{f},n^{i}) <n_{2c}^{f} \mid
t_{\pm}\sigma_{-\mu} \mid p_{c}^{i}>
\end{eqnarray}
while the four-q.p. states are simplified as
\begin{eqnarray}
<p_{1}^{f}p_{2}^{f}n_{1}^{f}n_{2c}^{f} \mid t_{\pm}\sigma_{-\mu}
\mid p_{1}^{i}p_{2}^{i}p_{3}^{i}n_{1c}^{i}>
\nonumber\\
=\delta (n_{2}^{f},n_{1}^{i})[ \delta (p_{1}^{f},p_{2}^{i})\delta
(p_{2}^{f},p_{3}^{i})
<n_{1c}^{f} \mid t_{\pm}\sigma_{-\mu} \mid p_{1c}^{i}> \nonumber\\
-\delta (p_{1}^{f},p_{1}^{i}) \delta (p_{2}^{f},p_{3}^{i})
<n_{1c}^{f} \mid t_{\pm}\sigma_{-\mu} \mid p_{2c}^{i}> +\delta
(p_{1}^{f},p_{1}^{i}) \delta (p_{2}^{f},p_{2}^{i}) \nonumber\\
<n_{1c}^{f} \mid t_{\pm}\sigma_{-\mu} \mid p_{3c}^{i}>]  -\delta
(n_{1}^{f},n_{1}^{i})[ \delta (p_{1}^{f},p_{2}^{i})\delta
(p_{2}^{f},p_{3}^{i})
<n_{2c}^{f} \mid t_{\pm}\sigma_{-\mu} \mid p_{1c}^{i}> \nonumber\\
-\delta (p_{1}^{f},p_{1}^{i}) \delta (p_{2}^{f},p_{3}^{i})
<n_{2c}^{f} \mid t_{\pm}\sigma_{-\mu} \mid p_{2c}^{i}> +\delta
(p_{1}^{f},p_{1}^{i}) \delta (p_{2}^{f},p_{2}^{i})\nonumber\\
<n_{2c}^{f}\mid t_{\pm}\sigma_{-\mu} \mid p_{3c}^{i}>]
\end{eqnarray}
\begin{eqnarray}
<p_{1}^{f}p_{2}^{f}p_{3}^{f}p_{4c}^{f} \mid t_{\pm}\sigma_{\mu} \mid
p_{1}^{i}p_{2}^{i}p_{3}^{i}n_{1c}^{i}>
\nonumber\\
=-\delta (p_{2}^{f},p_{1}^{i}) \delta (p_{3}^{f},p_{2}^{i})\delta
(p_{4}^{f},p_{3}^{i})
<p_{1c}^{f} \mid t_{\pm}\sigma_{\mu} \mid n_{1c}^{i}> \nonumber\\
+\delta (p_{1}^{f},p_{1}^{i}) \delta (p_{3}^{f},p_{2}^{i}) \delta
(p_{4}^{f},p_{3}^{i})
<p_{2c}^{f} \mid t_{\pm}\sigma_{\mu} \mid n_{1c}^{i}> \nonumber\\
-\delta (p_{1}^{f},p_{1}^{i}) \delta (p_{2}^{f},p_{2}^{i}) \delta
(p_{4}^{f},p_{3}^{i})
<p_{3c}^{f} \mid t_{\pm}\sigma_{\mu} \mid n_{1c}^{i}> \nonumber\\
+\delta (p_{1}^{f},p_{1}^{i}) \delta (p_{2}^{f},p_{2}^{i}) \delta
(p_{3}^{f},p_{3}^{i}) <p_{4c}^{f} \mid t_{\pm}\sigma_{\mu} \mid
n_{1c}^{i}>
\end{eqnarray}
\begin{eqnarray}
<p_{1}^{f}p_{2}^{f}n_{1}^{f}n_{2c}^{f} \mid t_{\pm}\sigma_{\mu} \mid
p_{1}^{i}n_{1}^{i}n_{2}^{i}n_{3c}^{i}>
\nonumber\\
=\delta (p_{1}^{f},p_{1}^{i})[ \delta (n_{1}^{f},n_{2}^{i})\delta
(n_{2}^{f},n_{3}^{i})
<p_{2c}^{f} \mid t_{\pm}\sigma_{\mu} \mid n_{1c}^{i}> \nonumber\\
-\delta (n_{1}^{f},n_{1}^{i}) \delta (n_{2}^{f},n_{3}^{i})
<p_{2c}^{f} \mid t_{\pm}\sigma_{\mu} \mid n_{2c}^{i}> +\delta
(n_{1}^{f},n_{1}^{i}) \delta (n_{2}^{f},n_{2}^{i})\nonumber\\
<p_{2c}^{f} \mid t_{\pm}\sigma_{\mu} \mid n_{3c}^{i}>]  -\delta
(p_{2}^{f},p_{1}^{i})[ \delta (n_{1}^{f},n_{2}^{i})\delta
(n_{2}^{f},n_{3}^{i})
<p_{1c}^{f} \mid t_{\pm}\sigma_{\mu} \mid n_{1c}^{i}> \nonumber\\
-\delta (n_{1}^{f},n_{1}^{i}) \delta (n_{2}^{f},n_{3}^{i})
<p_{1c}^{f} \mid t_{\pm}\sigma_{\mu} \mid n_{2c}^{i}> +\delta
(n_{1}^{f},n_{1}^{i}) \delta (n_{2}^{f},n_{2}^{i}) \nonumber\\
<p_{1c}^{f} \mid t_{\pm}\sigma_{\mu} \mid n_{3c}^{i}>]
\end{eqnarray}
\begin{eqnarray}
<n_{1}^{f}n_{2}^{f}n_{3}^{f}n_{4c}^{f} \mid t_{\pm}\sigma_{-\mu}
\mid p_{1}^{i}n_{1}^{i}n_{2}^{i}n_{3c}^{i}>
\nonumber\\
= +\delta (n_{2}^{f},n_{1}^{i}) \delta (n_{3}^{f},n_{2}^{i})\delta
(n_{4}^{f},n_{3}^{i})
<n_{1c}^{f} \mid t_{\pm}\sigma_{-\mu} \mid p_{1c}^{i}> \nonumber\\
-\delta (n_{1}^{f},n_{1}^{i}) \delta (n_{3}^{f},n_{2}^{i}) \delta
(n_{4}^{f},n_{3}^{i})
<n_{2c}^{f} \mid t_{\pm}\sigma_{-\mu} \mid p_{1c}^{i}> \nonumber\\
+\delta (n_{1}^{f},n_{1}^{i}) \delta (n_{2}^{f},n_{2}^{i}) \delta
(n_{4}^{f},n_{3}^{i})
<n_{3c}^{f} \mid t_{\pm}\sigma_{-\mu} \mid p_{1c}^{i}> \nonumber\\
-\delta (n_{1}^{f},n_{1}^{i}) \delta (n_{2}^{f},n_{2}^{i}) \delta
(n_{3}^{f},n_{3}^{i}) <n_{4c}^{f} \mid t_{\pm}\sigma_{-\mu} \mid
p_{1c}^{i}> \label{last}
\end{eqnarray}
For all the given q.p. transition amplitudes [Eqs. ~(\ref{first})- ~(\ref{last})],
the antisymmetrization of the single- q.p. states was taken into account:\\
$ p_{1}^{f}<p_{2}^{f}<p_{3}^{f}<p_{4}^{f}$,\\
$ n_{1}^{f}<n_{2}^{f}<n_{3}^{f}<n_{4}^{f}$,\\
$ p_{1}^{i}<p_{2}^{i}<p_{3}^{i}<p_{4}^{i}$,\\
$ n_{1}^{i}<n_{2}^{i}<n_{3}^{i}<n_{4}^{i}$.\\
GT transitions of phonon excitations for every excited state were
also taken into account. Here I assumed that the quasiparticles in
the parent nucleus remained in the same quasiparticle orbits. A
detailed description of the formalism can be found in Ref.
\cite{Mut92}.

\section{Results and discussion}
The improved pn-QRPA model, as described in the previous section,
was used to calculate the GT$_{\pm}$ strength distributions of iron
isotopes in astrophysical conditions. The model incorporated
experimental excitation energies, log$ft$ values and deformations as
discussed earlier. The GT strength distributions (in both
directions) were calculated for ground and 245 excited states in
$^{54}$Fe, 296 excited states in $^{55}$Fe and 265 excited states in
$^{56}$Fe. The model was tested for the case of $^{54,56}$Fe where
measured distributions were available. It was shown in Ref.
\cite{Nab09a} that the comparison of GT$_{\pm}$ strength
distributions for $^{54,56}$Fe with measured data was very good
(e.g. the total B(GT$_{-}$) strength for $^{54}$Fe calculated within
the framework of Ref. \cite{Nab04} was 9.33 which was narrowed down
to 7.56 using the improved model of pn-QRPA \cite{Nab09a} in
comparison with the measured value of 7.5 $\pm$ 0.7 \cite{And90}).
The GT$_{\pm}$ strength distributions for ground and all excited
states of iron isotopes are available as ASCII files and can be
requested from the author.

In order to highlight the important role of parent excited states in
the calculation of total decay rates, I present in Table~1 the
values of the total beta decay ($\lambda_{bd}$) and positron decay
($\lambda_{pd}$) rates with contributions from all excited states in
units of $s^{-1}$. These rates are calculated at selected values of
astrophysical temperature and density shown in first column (the
first number within the parenthesis gives the value of density in
units of $gcm^{-3}$ and the second number denotes the stellar
temperature in units of $10^{9}$ K). It is to be noted that the
corresponding ground-state beta and positron decay rates are
energetically forbidden. The excited states play a key role in the
calculation of total decay rates which are very sensitive to the
available phase space ($= Q_{\beta} +E_{i} - E_{j}$). The
construction of excited states in the pn-QRPA model was discussed in
the previous section. The phase space integrals given in
Eq.~(\ref{phase space}) increase considerably at higher stellar
temperatures and cause orders of magnitude enhancement in the
calculated decay rates. Fine grid calculation of positron and
$\beta$-decay rates for iron isotopes as a function of stellar
temperature, density and Fermi energy, suitable for core-collapse
simulations and interpolation purposes, is available as ASCII files
and can be requested from the author.

\begin{figure}[htbp]
\caption{(Color online) $\beta^{-}$ decay rates of $^{54}$Fe, as a
function of stellar temperatures, for different selected densities.
Densities are in units of $gcm^{-3}$. Temperatures are given in
$10^{9}$ K and log$_{10}\lambda_{\beta^{-}}$ represents the log to
base 10 of electron decay rates in units of $sec^{-1}$.
\label{fig1}}
\begin{center}
\begin{tabular}{c}
\includegraphics[width=0.8\textwidth]{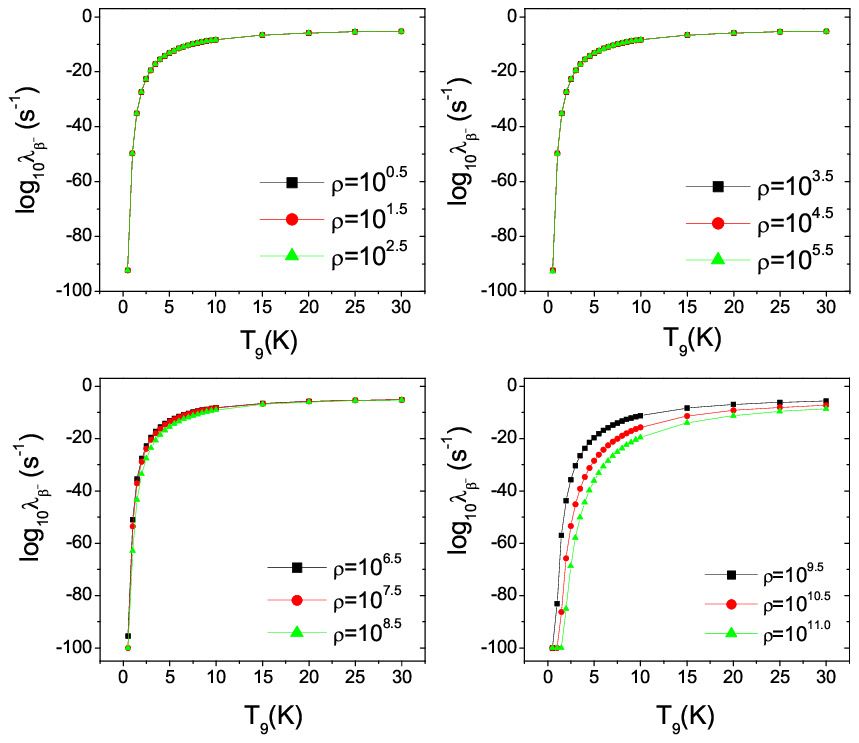}
\end{tabular}
\end{center}
\end{figure}

\begin{figure}[htbp]
\caption{(Color online) Same as figure~\ref{fig1} but for
$\beta^{-}$ decay rates of $^{55}$Fe. \label{fig2}}
\begin{center}
\begin{tabular}{c}
\includegraphics[width=0.8\textwidth]{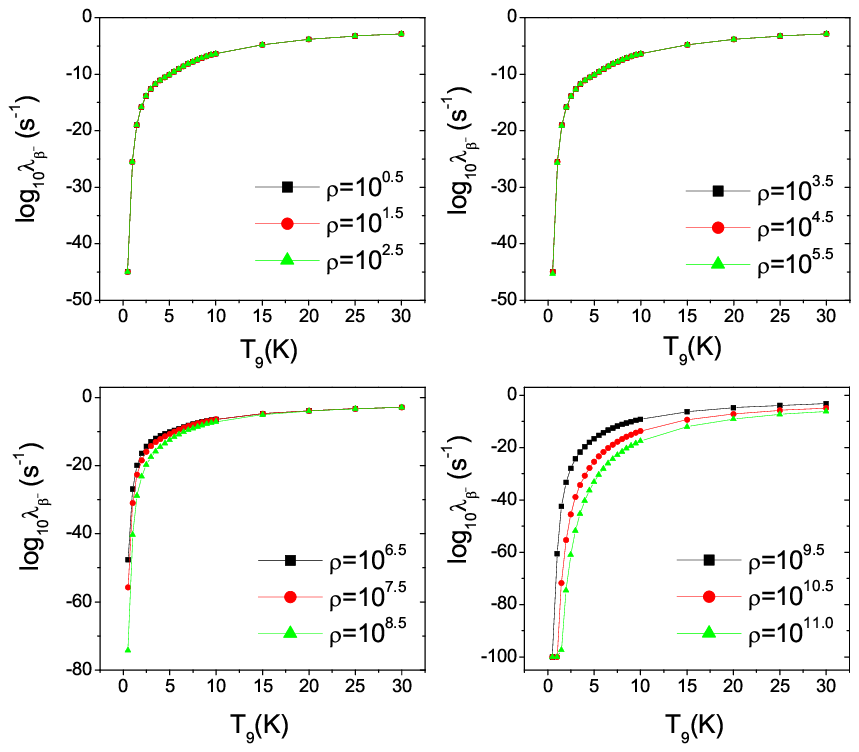}
\end{tabular}
\end{center}
\end{figure}

\begin{figure}[htbp]
\caption{(Color online) Same as figure~\ref{fig1} but for
$\beta^{-}$ decay rates of $^{56}$Fe. \label{fig3}}
\begin{center}
\begin{tabular}{c}
\includegraphics[width=0.8\textwidth]{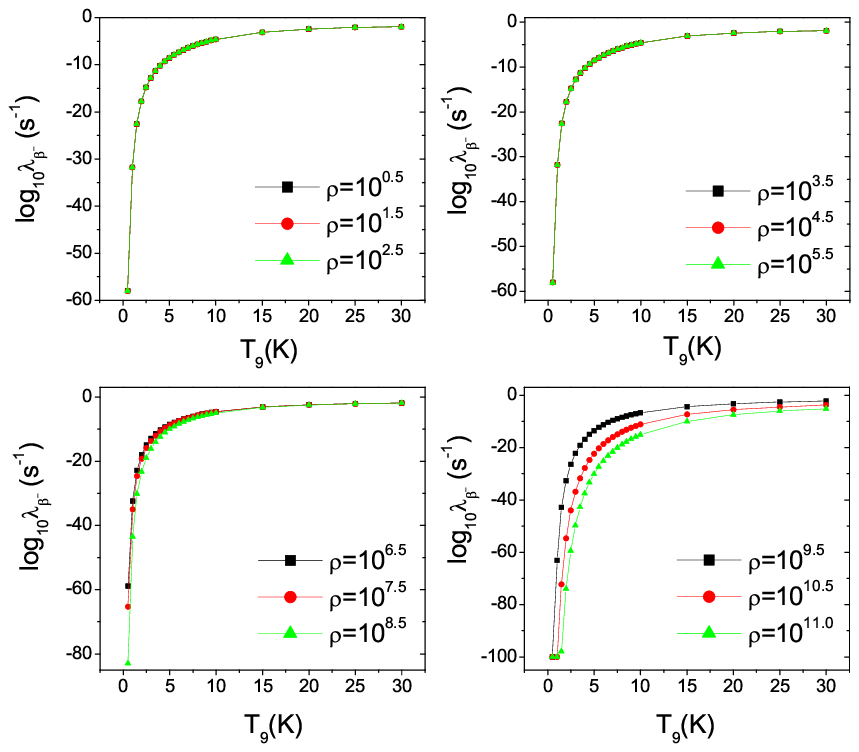}
\end{tabular}
\end{center}
\end{figure}

Figures~\ref{fig1}, \ref{fig2} and \ref{fig3} show the calculated
$\beta$-decay rates for $^{54,55,56}$Fe, respectively. Each figure
shows four panels depicting the calculated $\beta$-decay rates at
selected temperature and density domain. The upper left panel shows
the decay rates in low-density region ($\rho Y_{e} [gcm^{-3}]
=10^{0.5}, 10^{1.5}$ and $10^{2.5}$), the upper right in medium-low
density region ($\rho Y_{e} [gcm^{-3}] =10^{3.5}, 10^{4.5}$ and
$10^{5.5}$), the lower left in medium-high density region ($\rho
Y_{e} [gcm^{-3}] =10^{6.5}, 10^{7.5}$ and $10^{8.5}$) and finally
the lower right panel depicts the calculated rates in high density
region ($\rho Y_{e} [gcm^{-3}] =10^{9.5}, 10^{10.5}$ and $10^{11}$)
of stellar core of massive stars. The decay rates are given in
logarithmic scales (to base 10) in units of $s^{-1}$. T$_{9}$ gives
the stellar temperature in units of $10^{9}$ K. One should note the
orders of magnitude increment in $\beta$-decay rates as the stellar
temperature increases. The rates are almost superimposed on one
another as a function of stellar density in the first three panels.
This means that there is no appreciable change in the $\beta$-decay
rates when increasing the density by an order of magnitude. However
as the stellar matter moves from the medium high density region to
high density region these rates start to 'peel off' from one
another. In high density regions the rates start to decrease
appreciably due to a decrease in the phase space. There is a sharp
exponential increase in the decay rates as the stellar temperature
increases up to $\sim$ T$_{9}$ = 5. Beyond this temperature the
slope of the rates reduces drastically with increasing density. For
a given density the $\beta$-decay rates increase monotonically with
increasing temperatures. The calculated stellar $\beta$-decay rates
are smallest for $^{54}$Fe and biggest for $^{55}$Fe. The calculated
$\beta$-decay rates are smaller in magnitude compared to the
corresponding positron capture rates (in the same direction) on
these iron isotopes.

\begin{figure}[htbp]
\caption{(Color online) Same as figure~\ref{fig1} but for
$\beta^{+}$ decay rates of $^{54}$Fe. \label{fig4}}
\begin{center}
\begin{tabular}{c}
\includegraphics[width=0.8\textwidth]{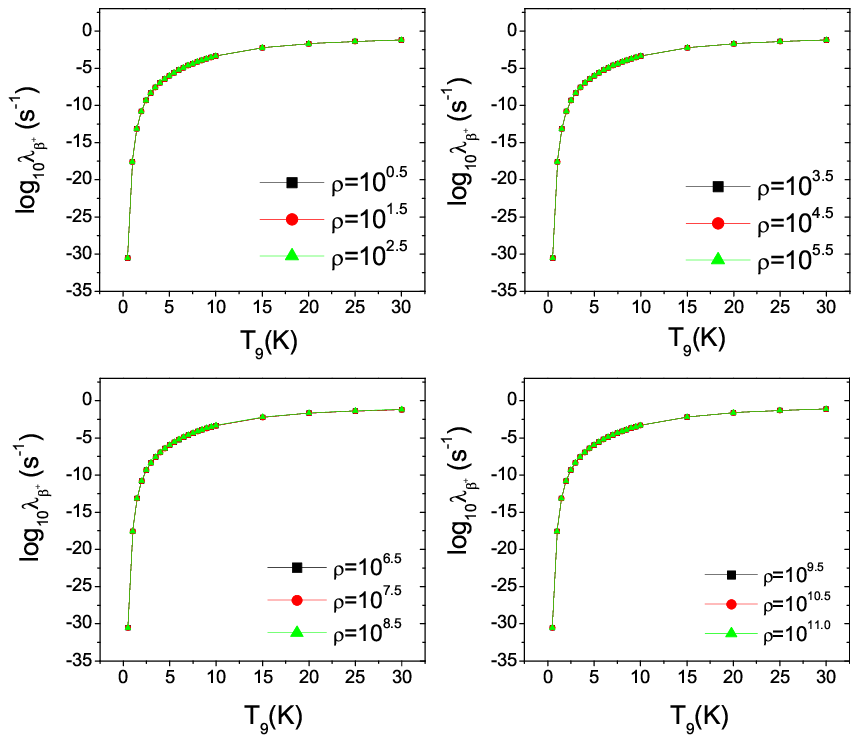}
\end{tabular}
\end{center}
\end{figure}

\begin{figure}[htbp]
\caption{(Color online) Same as figure~\ref{fig1} but for
$\beta^{+}$ decay rates of $^{55}$Fe. \label{fig5}}
\begin{center}
\begin{tabular}{c}
\includegraphics[width=0.8\textwidth]{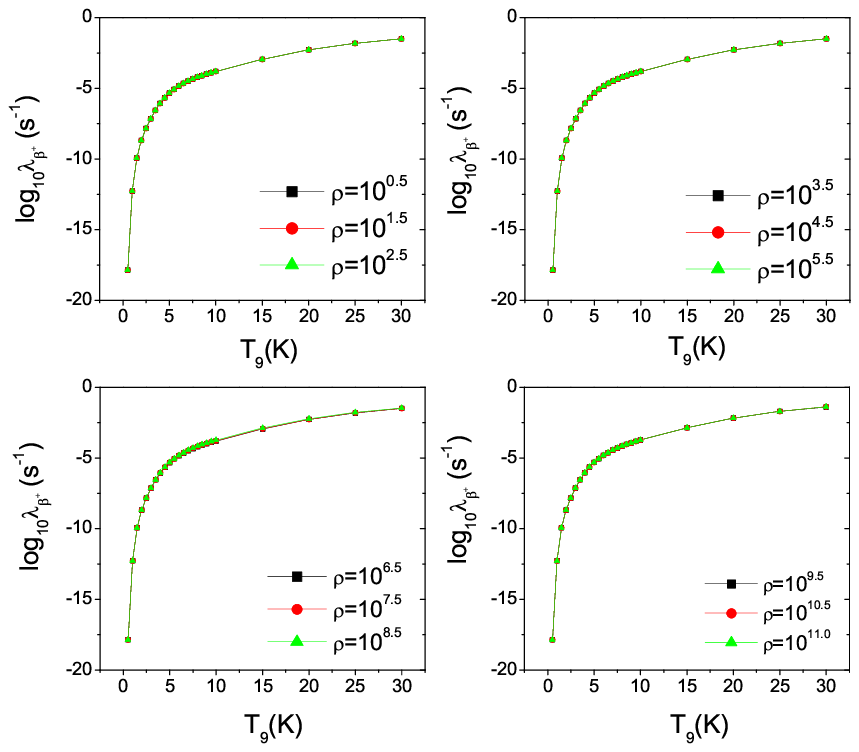}
\end{tabular}
\end{center}
\end{figure}

\begin{figure}[htbp]
\caption{(Color online) Same as figure~\ref{fig1} but for
$\beta^{+}$ decay rates of $^{56}$Fe. \label{fig6}}
\begin{center}
\begin{tabular}{c}
\includegraphics[width=0.8\textwidth]{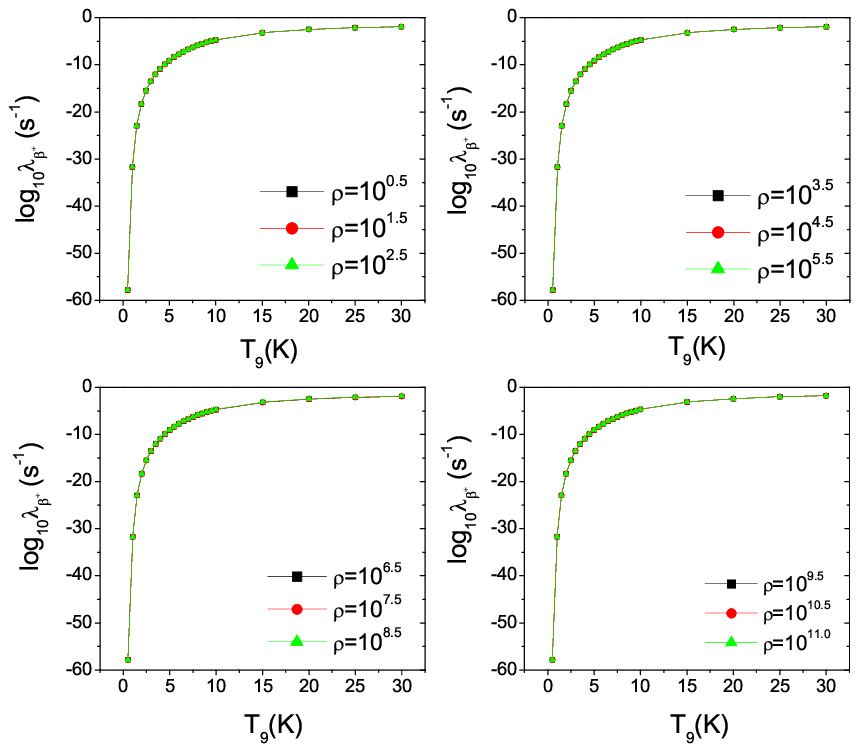}
\end{tabular}
\end{center}
\end{figure}

Figures~\ref{fig4}, \ref{fig5} and \ref{fig6} again show four panels
depicting the calculated positron decay rates at selected
temperature and density domain for $^{54,55,56}$Fe, respectively.
The upper left panel shows the positron decay rates in low-density
region ($\rho Y_{e} [gcm^{-3}] =10^{0.5}, 10^{1.5}$ and $10^{2.5}$),
the upper right in medium-low density region ($\rho Y_{e} [gcm^{-3}]
=10^{3.5}, 10^{4.5}$ and $10^{5.5}$), the lower left in medium-high
density region ($\rho Y_{e} [gcm^{-3}] =10^{6.5}, 10^{7.5}$ and
$10^{8.5}$) and finally the lower right panel depicts the calculated
positron decay rates in high density region ($\rho Y_{e} [gcm^{-3}]
=10^{9.5}, 10^{10.5}$ and $10^{11}$) of the stellar core. The rates
are given in logarithmic scales (to base 10) in units of $s^{-1}$.
The calculated positron decay rates are greater than the
corresponding $\beta$-decay rates by orders of magnitude. Due to an
increase in phase space with increasing temperature, the positron
decay rates increase by orders of magnitude as T$_{9}$ increases.
The increase is exponential up to $\sim$ T$_{9}$ = 5. As temperature
rises (the degeneracy parameter is negative for positrons), more and
more high-energy positrons are created leading in turn to higher
decay rates.  It can be seen from these figures that the positron
decay rates are almost the same as functions of the core density.
The positron decay rates are biggest for $^{55}$Fe and smallest for
$^{56}$Fe. It is worth mentioning that the beta and positron decay
rates are very small numbers and can change by orders of magnitude
by a mere change of 0.5 MeV, or less, in parent or daughter
excitation energies (as the rates are very sensitive to the
available Q$_{\beta}$ window) and are more reflective of the
uncertainties in the calculation of energies.

It would be interesting to know how the reported decay rates of iron
isotopes compare with earlier calculations for temperature and
density domains of astrophysical interest. For the sake of
comparison I took into consideration the pioneer calculations of FFN
\cite{Ful80} and those performed using the large-scale shell model
(LSSM) \cite{Lan00}. The FFN rates were used in many simulation
codes (e.g. KEPLER stellar evolution code \cite{Wea78}) while LSSM
rates were employed in recent simulation of presupernova evolution
of massive stars in the mass range 11-40 $M_{\odot}$ \cite{Heg01}.

\begin{figure}[htbp]
\caption{(Color online) Ratios of reported $\beta^{+}$ decay rates
to those calculated using LSSM \cite{Lan00} (upper panel) and FFN
\cite{Ful80} (lower panel) for $^{54}$Fe as function of stellar
temperatures and densities. T$_{9}$ gives the stellar temperature in
units of $10^{9}$ K. In the legend, $log \rho Y_{e}$ gives the log
to base 10 of stellar density in units of $gcm^{-3}$. \label{fig7}}
\begin{center}
\begin{tabular}{c}
\includegraphics[width=0.8\textwidth]{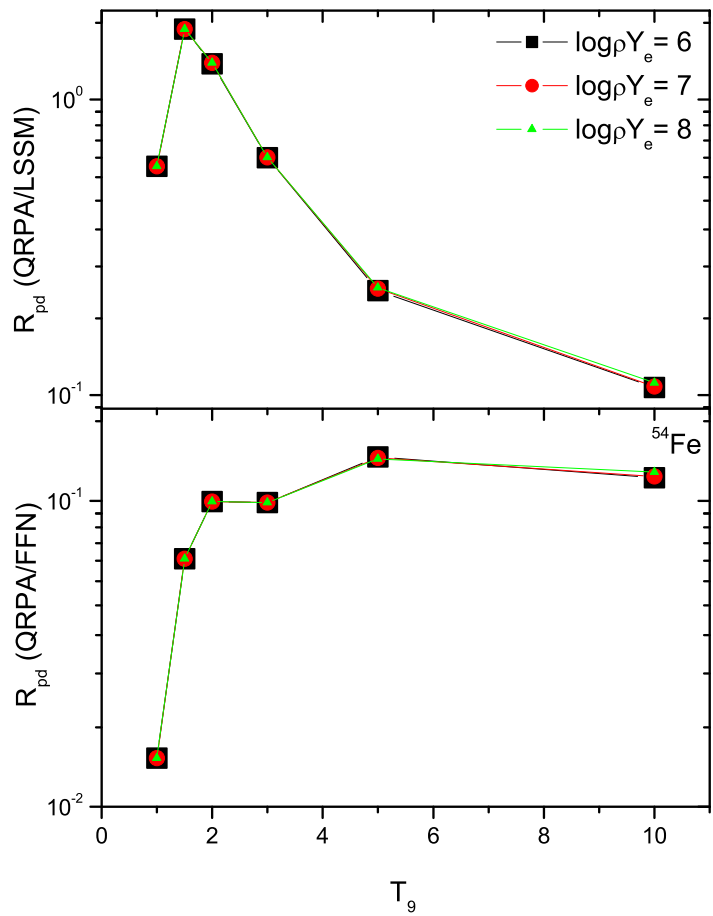}
\end{tabular}
\end{center}
\end{figure}

Positron decay rates of iron isotopes may decrease the $Y_{e}$ value
of massive stars.  As mentioned before, the positron decay rate
calculations do not differ appreciably by changing densities.
Figure~\ref{fig7} depicts the comparison of positron decay rates for
$^{54}$Fe with earlier calculations. The upper panel displays the
ratio of calculated rates to the LSSM rates, $R_{pd}(QRPA/LSSM)$,
while the lower panel shows a similar comparison with the FFN
calculation, $R_{pd}(QRPA/FFN)$. The comparison is made for the
selected temperature and density domain. Here one sees that the
pn-QRPA and LSSM calculations are in overall reasonable agreement.
The comparison is good at low temperatures while at higher
temperatures, $T_{9} > $ 5, the LSSM rates are bigger around an
order of magnitude. The FFN rates are much bigger than the reported
positron decay rates by  1 -- 2 orders of magnitude. It is reminded
that FFN neglected the quenching of the GT strength in their rate
calculation. Further FFN did not take into effect the process of
particle emission from excited states and their parent excitation
energies extended well beyond the particle decay channel. These high
lying excited states began to show their cumulative effect at high
temperatures and densities.

\begin{figure}[htbp]
\caption{(Color online) Same as figure~\ref{fig7} but for
$\beta^{+}$ decay rates of $^{55}$Fe. \label{fig8}}
\begin{center}
\begin{tabular}{c}
\includegraphics[width=0.8\textwidth]{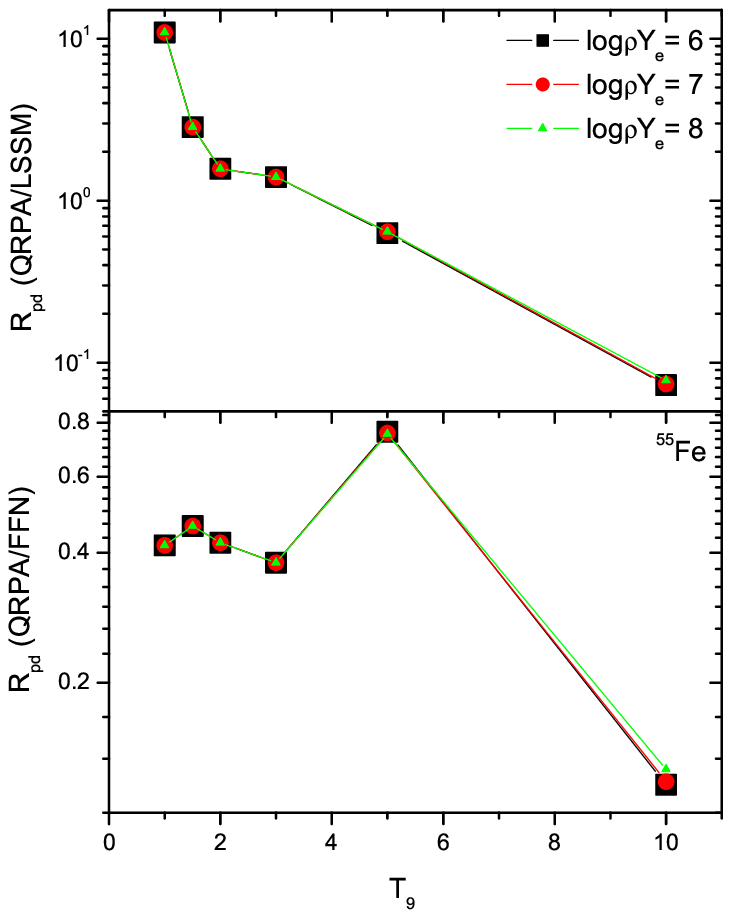}
\end{tabular}
\end{center}
\end{figure}

The comparison of reported positron decay rates of $^{55}$Fe with
LSSM calculation is somewhat different. For the first time one sees
that the pn-QRPA calculated positron decay rate is bigger than LSSM
(Figure~\ref{fig8}). This however is true only for $T_{9}$ = 1.
During the early phases of presupernova evolution the reported
positron decay rates are bigger than LSSM rates by an order of
magnitude. For successive stages, 1.5 $\leq T_{9} \leq $ 5, the two
calculations are in good agreement. At still higher temperatures the
LSSM rates exceed by more than an order of magnitude. Excited state
GT distributions in $^{55}$Fe play a key role in determining the
total decay rate at various stages of presupernova evolution. It is
reminded that the pn-QRPA makes a microscopic assessment of GT
strength distribution for all such parent excited states. Comparison
with the FFN rates is fairly constant. It is to be noted that the
ratios are shown on a linear scale (bottom panel of
Figure~\ref{fig8}). At $T_{9}$ = 10, the FFN rates surpass the
reported rates for reasons already mentioned.

\begin{figure}[htbp]
\caption{(Color online) Same as figure~\ref{fig7} but for
$\beta^{+}$ decay rates of $^{56}$Fe. \label{fig9}}
\begin{center}
\begin{tabular}{c}
\includegraphics[width=0.8\textwidth]{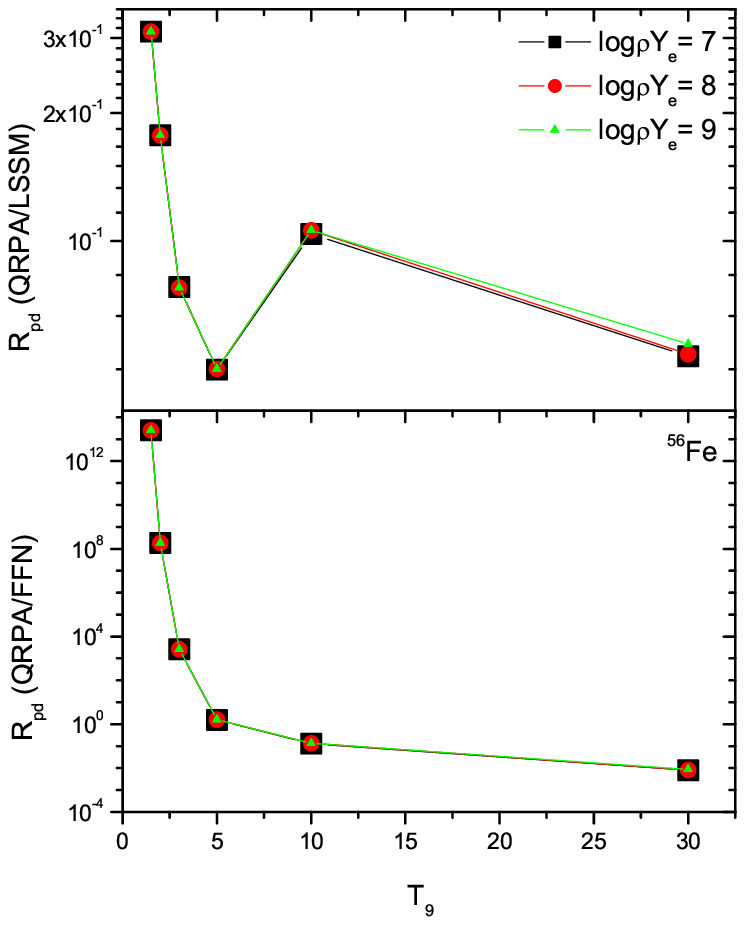}
\end{tabular}
\end{center}
\end{figure}

The comparison of reported positron decay rates of  $^{56}$Fe with
previous calculations is shown in Figure~\ref{fig9}. The LSSM decay
rates are bigger by as much as a factor of 20 (the comparison is
better at low temperatures).  The lower panel of Figure~\ref{fig9}
shows a whopping enhancement in the calculated positron decay rates
of $^{56}$Fe of around 13 orders of magnitude compared to FFN rates
at low temperatures. However there are reasons for this unusual
suppression of FFN positron decay rates. Unmeasured matrix elements
for allowed transitions were assigned an average value of $log ft =
$5 in FFN calculations. Later, $(n,p)$ experiments were performed
for $^{56}$Fe \cite{Roe93, Elk94} which revealed a too small $log
ft$ value assignment for allowed transitions employed by FFN.
Consequently the FFN rates are much too smaller than the LSSM and
pn-QRPA calculations at low temperatures (LSSM rates are also bigger
than FFN rates by roughly 13 orders of magnitude at $T_{9}$ = 1.5).
The LSSM positron decay rates for iron isotopes are bigger up to a
factor of 20 as compared to the pn-QRPA rates. The overall
comparison of positron decay rates of $^{54,55,56}$Fe shows that the
calculated rates are in reasonable agreement with the LSSM
calculated rates. Major differences are seen in the case of the
$\beta$-decay rate calculations and are discussed in detail below.

Authors in Ref. \cite{Lan00} reported that, for even-even and odd-A
nuclei, FFN systematically placed the back resonance at much lower
energies and concluded that contribution of the back resonances to
the $\beta$-decay rates for these nuclei decreases. They estimated
that LSSM $\beta$-decay rates as a result were smaller, on the
average, by a factor of 20 (40) as compared to the FFN $\beta$-decay
rates for even-even (odd-A) nuclei.

\begin{figure}[htbp]
\caption{(Color online) Same as figure~\ref{fig7} but for
$\beta^{-}$ decay rates of $^{54}$Fe. \label{fig10}}
\begin{center}
\begin{tabular}{c}
\includegraphics[width=0.8\textwidth]{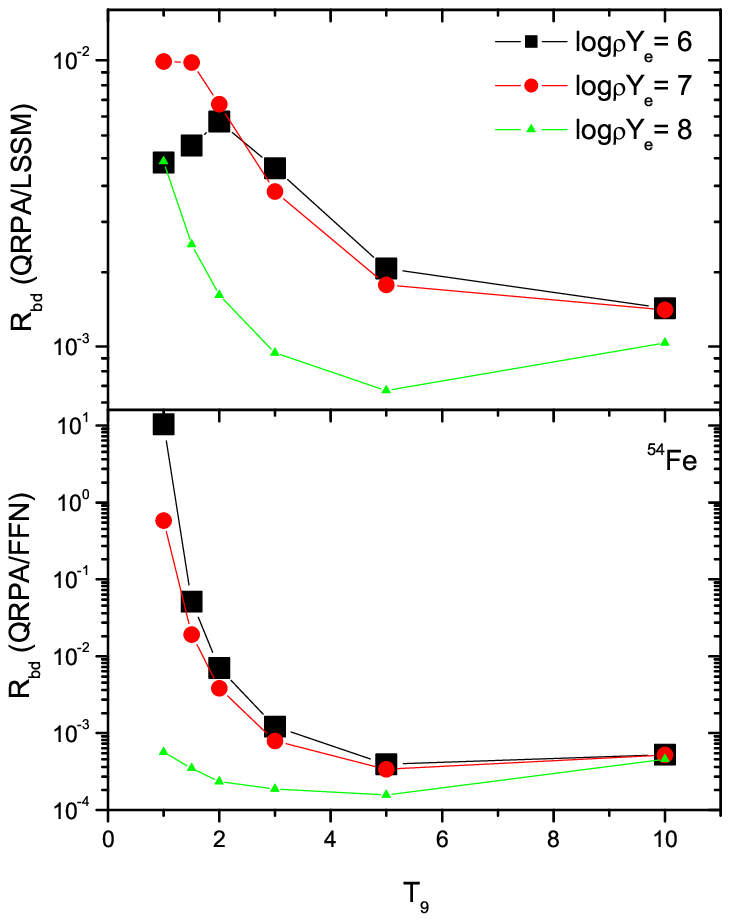}
\end{tabular}
\end{center}
\end{figure}

Figure~\ref{fig10} depicts the comparison of $\beta$-decay rates for
$^{54}$Fe with earlier calculations. As before the upper panel
displays the ratio of calculated rates to the LSSM rates,
$R_{bd}(QRPA/LSSM)$, while the lower panel shows a similar
comparison with the FFN calculation, $R_{bd}(QRPA/FFN)$. The density
scale is shown in the inset. A mutual comparison of LSSM and FFN
$\beta$-decay rates for the case of $^{54}$Fe reveals that for low
temperatures (T$_{9} < $ 2) and densities $\rho Y_{e} [gcm^{-3}]
\sim 10^{6}-10^{7}$ the LSSM $\beta$-decay rates are bigger than the
FFN rates by as much as four orders of magnitude. Only at higher
temperatures and densities are the LSSM rates smaller than the FFN
rates (by around an order of magnitude). The LSSM $\beta$-decay
rates for $^{54}$Fe are 3 -- 4 orders of magnitude bigger than the
pn-QRPA rates for the physical conditions depicted in
Figure~\ref{fig10}. The enhancement ratio increases with increasing
densities. The reasons for bigger LSSM $\beta$-decay rates are not
very clear. The Q-values and phase space formulation for the two
calculations appear the same. Perhaps the contribution from back
resonances needs a further cut back in the LSSM calculation for the
$\beta$-decay of even-even and odd-A nuclei (i.e. the back
resonances for these nuclei should be put at even higher energies
than calculated by LSSM). As mentioned before the reported rates do
not employ these approximations and calculate all the excited state
GT strength distributions in a microscopic fashion. The comparison
with the FFN rates reveals that the reported decay rates are smaller
by as much as four orders of magnitude (at higher temperatures). At
temperature T$_{9}$ = 1 and density $\rho Y_{e} [gcm^{-3}] =
10^{6}$, where the LSSM $\beta$-decay rates are bigger than the FFN
rates by roughly four orders of magnitude, the reported rates are
bigger than the FFN rates by an order of magnitude. For all other
points the FFN rates are much bigger than pn-QRPA rates for reasons
mentioned before.

\begin{figure}[htbp]
\caption{(Color online) Same as figure~\ref{fig7} but for
$\beta^{-}$ decay rates of $^{55}$Fe. \label{fig11}}
\begin{center}
\begin{tabular}{c}
\includegraphics[width=0.8\textwidth]{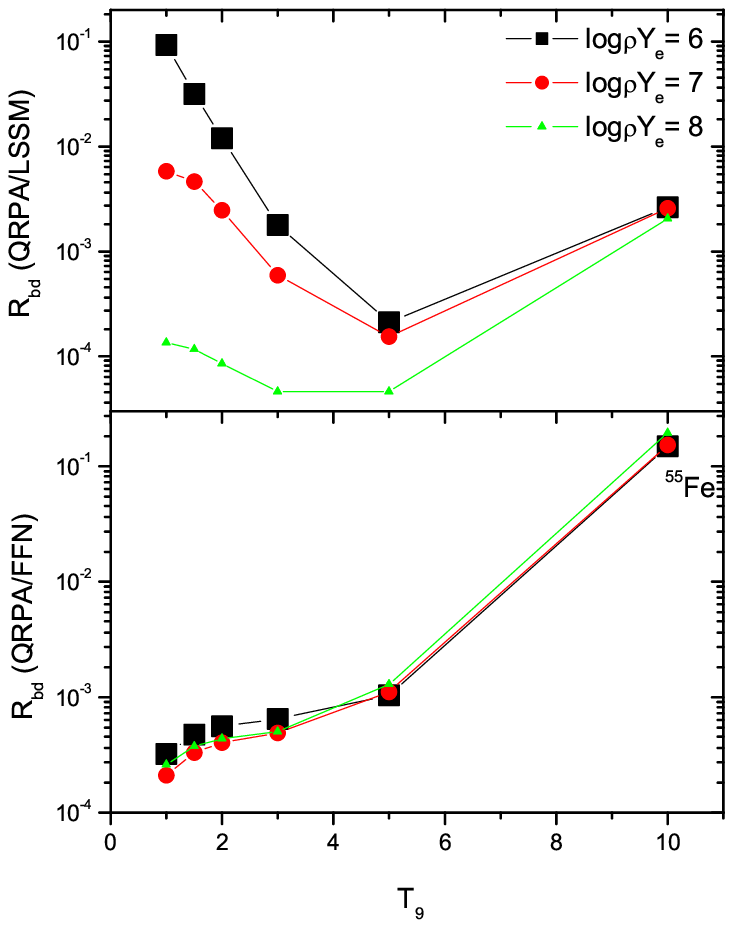}
\end{tabular}
\end{center}
\end{figure}

The $\beta$-decay rates of $^{55}$Fe are believed to be important
during the silicon burning stages of massive stars as per the
simulation results of Ref. \cite{Heg01}. Figure~\ref{fig11} shows a
comparison of the three calculations for $\beta$-decay rates of
$^{55}$Fe during the relevant temperature and density domain of
stellar core. Only at low temperatures and densities are the LSSM
decay rates smaller than the corresponding FFN numbers. At higher
temperatures (T$_{9} \ge $ 5) and densities ($\rho Y_{e} [gcm^{-3}]
> 10^{7}$) the LSSM rates surpass the FFN decay rates  by more than
two orders of magnitude. The comparison of pn-QRPA rates with LSSM
suggests that the LSSM rates are bigger by 2 -- 3 orders of
magnitude (at lower densities the comparison is relatively better) .
In fact the LSSM $\beta$-decay rates are even bigger than their
calculated positron capture rates (only at higher temperatures their
positron capture rates surpass the $\beta$-decay rates). In contrast
the pn-QRPA calculated $\beta$-decay rates are suppressed as
compared to the corresponding positron capture rates for all
temperature and density scales. At higher densities the LSSM
$\beta$-decay rates are bigger by 3 -- 5 orders of magnitude. The
FFN rates are up to four orders of magnitude bigger than the pn-QRPA
rates for reasons mentioned before.

\begin{figure}[htbp]
\caption{(Color online) Same as figure~\ref{fig7} but for
$\beta^{-}$ decay rates of $^{56}$Fe. \label{fig12}}
\begin{center}
\begin{tabular}{c}
\includegraphics[width=0.8\textwidth]{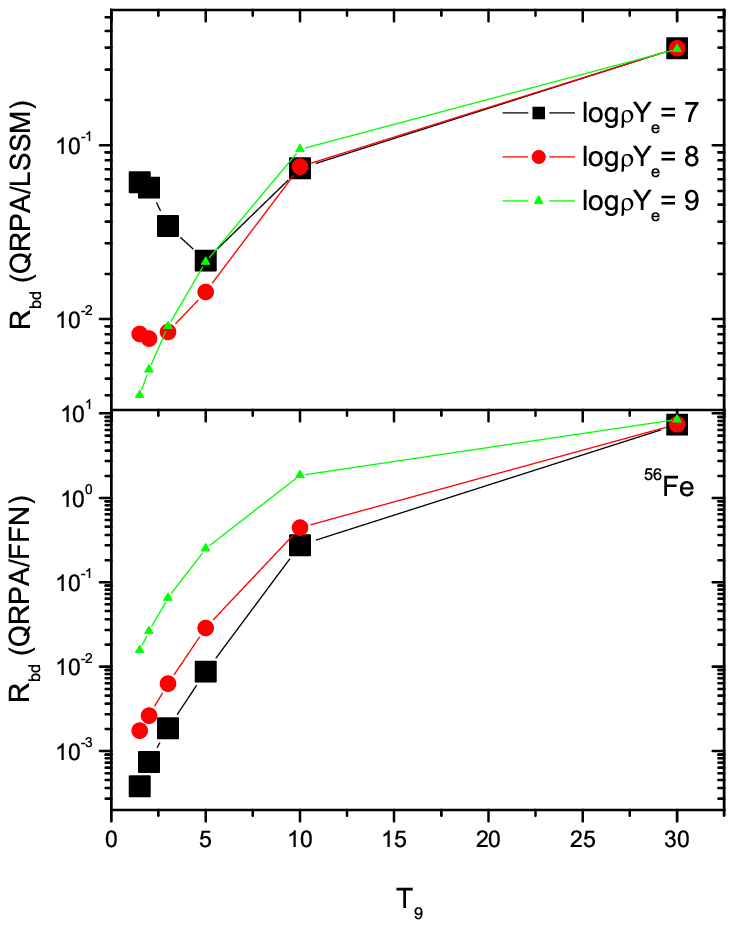}
\end{tabular}
\end{center}
\end{figure}

A study of LSSM and the FFN rates reveals that LSSM $\beta$-decay
rates of $^{56}$Fe are much bigger than the FFN rates at high
temperatures and densities (by more than an order of magnitude).
Figure~\ref{fig12} shows how the reported $\beta$-decay rates of
$^{56}$Fe compare with earlier calculations for relevant physical
conditions. It can be seen from the figure that the reported rates
are much smaller (up to four orders of magnitude) than previous
calculations. Comparison with LSSM calculation shows that at higher
densities the reported $\beta$-decay rates are suppressed by up to
three orders of magnitude. At higher temperatures (T$_{9}$ = 30) the
comparison ratio improves. However the LSSM rates are still bigger
by around an order of magnitude. On the other hand the FFN rates are
bigger than the reported $\beta$-decay rates by around four orders
of magnitude at lower densities and temperatures. The comparison
ratio improves with increasing temperatures and densities. The
pn-QRPA rates are bigger than the FFN rates by around a factor of 8
at T$_{9}$ = 30. The comparison study of stellar $\beta$-decay rates
of iron isotopes suggests that the LSSM $\beta$-decay rates are
bigger than the corresponding pn-QRPA rates by 3 -- 5 orders of
magnitude.

\begin{figure}[htbp]
\caption{Excited state Gamow-Teller (GT$_{-}$) strength
distributions for $^{54}$Fe.  $E_{i}$($E_{j}$) represents parent
(daughter) energy states. \label{fig13}}
\begin{center}
\begin{tabular}{c}
\includegraphics[width=0.8\textwidth]{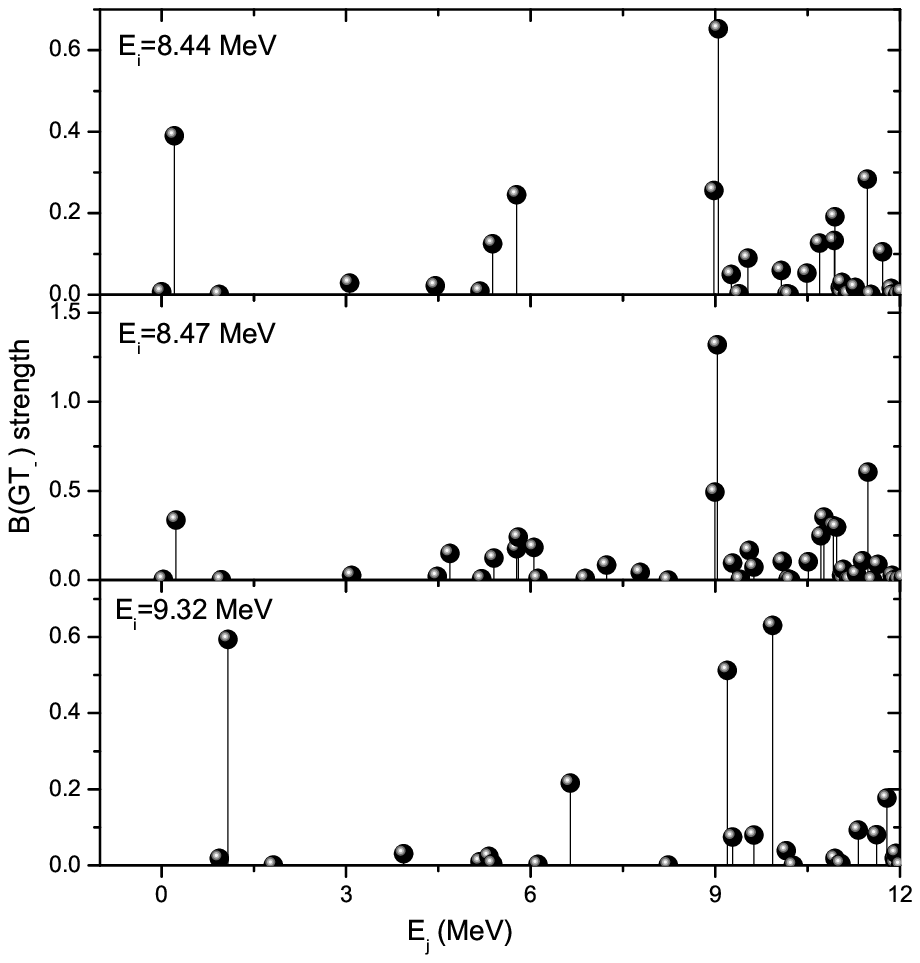}
\end{tabular}
\end{center}
\end{figure}

\begin{figure}[htbp]
\caption{Excited state Gamow-Teller (GT$_{-}$) strength
distributions for $^{55}$Fe.  $E_{i}$($E_{j}$) represents parent
(daughter) energy states. \label{fig14}}
\begin{center}
\begin{tabular}{c}
\includegraphics[width=0.8\textwidth]{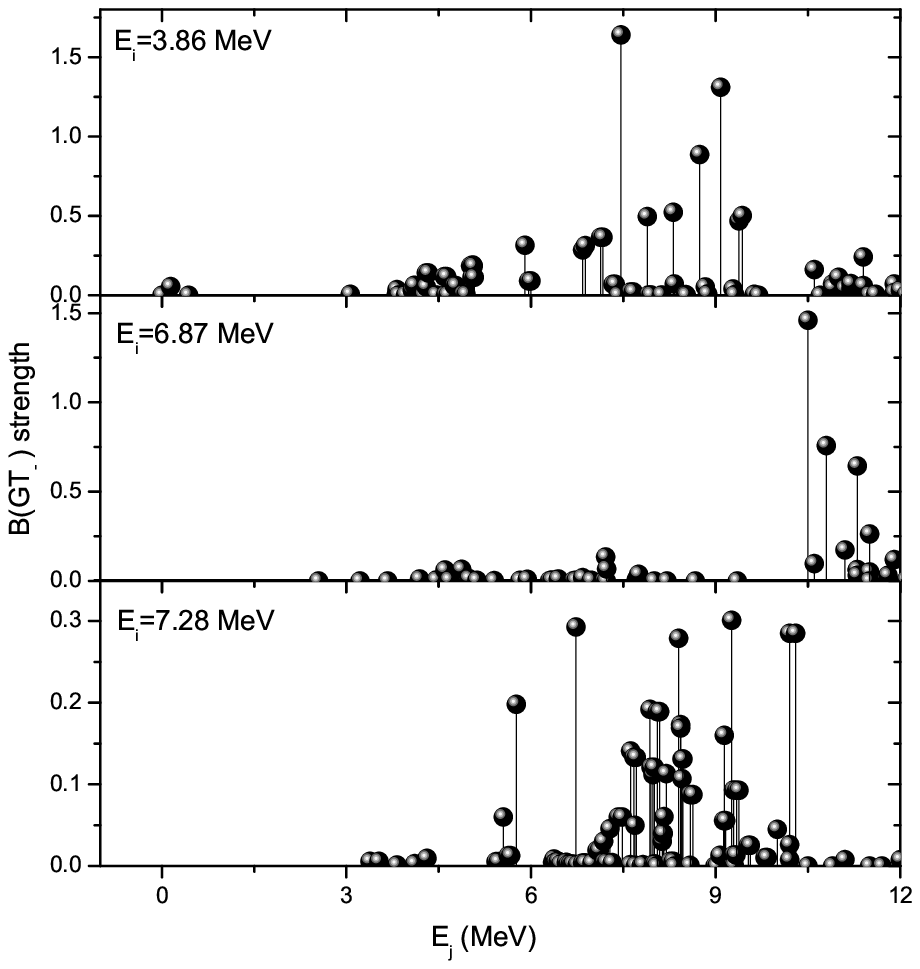}
\end{tabular}
\end{center}
\end{figure}

\begin{figure}[htbp]
\caption{Excited state Gamow-Teller (GT$_{-}$) strength
distributions for $^{56}$Fe.  $E_{i}$($E_{j}$) represents parent
(daughter) energy states. \label{fig15}}
\begin{center}
\begin{tabular}{c}
\includegraphics[width=0.8\textwidth]{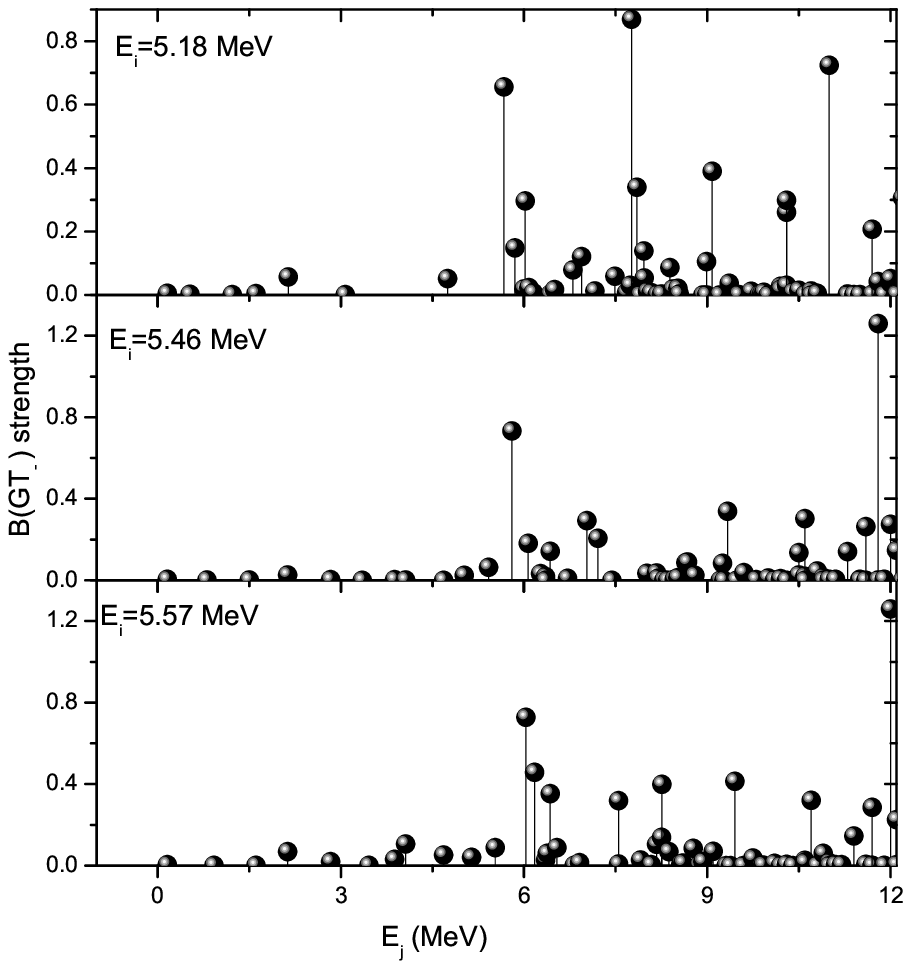}
\end{tabular}
\end{center}
\end{figure}

Earlier using the same model the electron capture rates due to
$^{54,55,56}$Fe were calculated and were found to be in overall good
comparison with the LSSM electron capture rates \cite{Nab09a}. In
fact the pn-QRPA electron capture rates on $^{54}$Fe were around a
factor three bigger than LSSM rates in the relevant astrophysical
conditions (see Table~3 of Ref. \cite{Nab09a}). Using the same
model, however, the calculated $\beta$-decay rates are smaller by 3
-- 5 orders of magnitude. This suppression in the pn-QRPA
$\beta$-decay rates comes from the excited state GT distributions
(smaller nuclear matrix elements) which are much different from the
ground state distribution. In order to gain a detailed insight of
these distributions I took a representative temperature and density
point of T$_{9}$ = 10 and $\rho Y_{e} [gcm^{-3}] = 10^{7}$ (weak
interaction rates due to iron isotopes are argued to contribute
effectively around such physical conditions during the presupernova
evolution of massive stars). Table~2 shows the data of the ground
and first four excited states that have a finite partial decay rate
contribution to the total $\beta$-decay rate of $^{54,55,56}$Fe
within the framework of the model of magnitude greater than
$10^{-25} s^{-1}$. The first column shows the calculated parent excited
energy state in units of MeV, the second column gives the product of
occupation probability and partial $\beta$-decay rate from this
state in units of $s^{-1}$. The third and fourth column give
the centroid and total B(GT) strength, respectively, from this parent state. The cut-off
energy in daughter nuclei is 12 MeV. It can be seen clearly from
Table~2 that for the even-even isotopes the centroids of the excited
state GT strength distributions are shifted to much higher energies
in the daughter. The $\Sigma S_{\beta^{-}}$ strengths are also
considerably smaller from the corresponding ground state strengths.
These are mainly responsible for the smaller total $\beta$-decay
rates. The excited state GT strength distributions are also much
different from the ground state distribution for the case of
$^{55}$Fe. The excited state GT distributions for $^{54,55,56}$Fe
are shown graphically in Figures~\ref{fig13}, \ref{fig14} and
\ref{fig15}, respectively. The ground state GT strength
distributions were presented earlier in Ref. \cite{Nab09a}. Here I
have shown only the first three excited state distributions that
have a finite partial decay rate contribution of magnitude greater than
$10^{-25} s^{-1}$. Note the different
scales of the B(GT) strength values in the vertical panels of these
figures. From these figures it is clear that the Brink's hypothesis
(and back resonances for the calculation of $\beta$-decay) is not a
good approximation for calculation of stellar weak interaction rates
of iron isotopes. These and similar finite partial decay rates lead
to an overall suppression in the total $\beta$-decay rate which
happens to be 5.44E-09 $s^{-1}$, 4.32E-07 $s^{-1}$ and 2.62E-05
$s^{-1}$ for $^{54}$Fe, $^{55}$Fe and $^{56}$Fe, respectively, under
the physical conditions stated above.

\section{Summary and conclusions}
Capture and decay rates are considered to be very important in
controlling the lepton-to-baryon ratio and entropy of the core of
massive stars during the presupernova evolutionary phases. These are
the two key parameters which later play a key role in the dynamics
of supernova explosion. A reliable and microscopic calculation of
these weak interaction rates can assist us in a better understanding
of the explosion process.

The pn-QRPA model has a very good rapport in calculation of beta
decay rates. The model has access to a huge model space of seven
major shells and is the only available model to perform a fully
microscopic 'state-by-state' calculation of weak rates in stellar
interior. The stellar electron and positron $\beta$-decay rates of
iron isotopes were presented using the pn-QRPA model. Incorporation
of latest experimental data and an optimum selection of model
parameters increased the reliability of the calculated rates. The
calculated rates were also compared with previous calculations.

The key findings of this work include that the Brink's hypothesis
and back resonances used in previous calculations of decay rates for
iron isotopes are not a good approximation to use. A microscopic
calculation of excited state GT strength distributions greatly
increases the reliability of calculated rates. The $\beta$-decay
rates calculated by large scale shell model calculation are 3 -- 5
orders of magnitude bigger than the reported rates in astrophysical
conditions. The microscopic calculation of excited state GT strength
distribution calculated within the framework of the pn-QRPA model,
which are responsible for the reduced $\beta$-decay rates, was also
discussed. During the early phases of presupernova evolution the
LSSM calculated $\beta$-decay rates are even bigger than the
competing positron capture rates (both occur in the same direction
and tend to decrease $Y_{e}$). The pn-QRPA calculated $\beta$-decay
rates are smaller than the competing positron capture rates. FFN
$\beta$-decay rates are up to four orders of magnitude bigger than
the reported rates under the same conditions.  The pn-QRPA
calculation validates the finding by authors in Ref. \cite{Lan00}
that FFN places the so-called back resonances at too low excitation
energies in even-even and odd-A nuclei. The reported calculation
also suggests that the placement of back resonances employed by LSSM
needs further push towards higher excitation energies. The LSSM
positron decay rates are, comparatively, in reasonable agreement
with the pn-QRPA rates.

What may be the astrophysical implications of the reported decay
rates of iron isotopes? The $\beta$-decay rates of iron nuclei are
much smaller than previously assumed and this news is important for
core-collapse simulators world-wide.  The current study suggests
that $\beta$-decay rates of iron isotopes are irrelevant for the
determination of the evolution of $Y_{e}$ during the presupernova
phase of massive stars. A review of inclusion of iron isotopes
$^{55,56}$Fe in the list of key stellar $\beta$-decay nuclei as
suggested by previous simulation results \cite{Auf94, Heg01} is in
order.

\vspace{0.5 in}\textbf{Acknowledgments:} The author would like to
acknowledge the kind hospitality provided by the Abdus Salam ICTP,
Trieste, where part of this project was completed. The author also
wishes to acknowledge the support of research grant provided by the
Higher Education Commission Pakistan,  through the HEC Project Nos.
20-1171 and 20-1283.


\include{table}
\include{tableA}

\end{document}